\def\ref{\par\noindent\hang}

\def\etal{{et al.\thinspace}}

\def\eg{{\em e.g.\ }}

\def\ie{{\em i.e.\ }}
\def\spose#1{\hbox to 0pt{#1\hss}}
\def\approxlt{\mathrel{\spose{\lower 3pt\hbox{$\sim$}}
        \raise 2.0pt\hbox{$$<$$}}}
\def\approxgt{\mathrel{\spose{\lower 3pt\hbox{$\sim$}}
        \raise 2.0pt\hbox{$>$}}}

\def\multleft#1{\hbox to size{\vbox {\halign {\lft{##}\cr #1}}\hfill}\par}
\def\multright#1{\hbox to size{\vbox {\halign {\rt{##}\cr #1}}\hfill}\par}

\def\today{\ifcase\month\or January\or February\or March\or April\or May\or
      June\or July\or August\or September\or October\or November\or December\fi
      \space\number\day, \number\year}
\def\$<${\thinspace}
\def\s{\hbox{\phantom{5}}}      %one space
\def\boxit#1{\vbox{\hrule\hbox{\vrule\kern3pt\vbox{\kern3pt
          #1 \kern3pt}\kern3pt\vrule}\hrule}}

\def\cm{{\rm\thinspace cm}}

\def\g{{\rm\thinspace g}}
\def\ga{{\rm\thinspace gauss}}
\def\K{{\rm\thinspace K}}

\def\km{{\rm\thinspace km}}

\def\Lsun{\hbox{$\rm\thinspace L_{\odot}$}}

\def\Mpc{{\rm\thinspace Mpc}}
\def\Msun{\hbox{$\rm\thinspace M_{\odot}$}}
\def\pc{{\rm\thinspace pc}}

\def\s{{\rm\thinspace s}}
\def\yr{{\rm\thinspace yr}}

\def\kmps{\hbox{$\km\s^{-1}\,$}}

\def\Msunpyr{\hbox{$\Msun\yr^{-1}\,$}}

\def\mic{{\rm\thinspace $\mu$m}}

\def\mm{{\rm\thinspace mm}}
\def\Gyr{{\rm\thinspace Gyr}}

\documentstyle[epsfig,psfig]{mn}
\begin{document}
\hsize=6truein

\title[The submm spectral energy distribution of high-redshift quasars]
{The far-infrared--submm spectral energy distribution of high-redshift quasars}
\author[R.S. Priddey \& R.G. McMahon]
{\parbox[]{6.5in}
{Robert S. Priddey$^{1,2}$ \& Richard G. McMahon$^1$}\\ 
{$^1$ \it Institute of Astronomy, Madingley Road, Cambridge CB3 0HA, UK}\\
{$^2$ \it Astrophysics Group, Blackett Laboratory, Imperial College, London SW7 2BZ, UK}\\
email: r.priddey@ic.ac.uk, rgm@ast.cam.ac.uk
}

\date{Accepted 3rd April 2001. Received 6th March 2001; 
in original form 18th October 2000}
%\date{Submitted to MNRAS}

\maketitle

\begin{abstract}
We combine photometric observations of high-redshift ($z>4$) quasars,
obtained at submillimetre (submm) to millimetre wavelengths,
to obtain a mean far-infrared (FIR) (rest-frame) 
spectral energy distribution (SED)
of the thermal emission from dust, parameterised by a single
temperature ($T$) and power-law emissivity index ($\beta$).
Best-fit values are $T=41\pm5\K$, $\beta=1.95\pm0.3$.
Our method exploits the redshift spread of this set of quasars,
which allows us to sample the SED at a larger number of 
rest wavelengths than is possible for a single object:
the wavelength range extends down to $\sim60$\mic, and
therefore samples the turnover in the greybody curve for 
these temperatures.
This parameterisation is of use to
any studies that extrapolate from a flux at a single wavelength,
for example to infer dust masses and FIR luminosities.

We interpret the cool, submm component as arising from dust
heated by star-formation in the quasar's host galaxy, although
we do not exclude the presence of dust
heated directly by the AGN.
Applying the mean SED to the data, we derive consistent
star-formation rates $\sim$1000\Msunpyr and dust masses 
$\sim10^9\Msun$, and investigate a simple scheme of AGN and 
host-galaxy coevolution to account for these quantities.
The timescale for formation of the host galaxy is 0.5--1Gyr,
and the luminous quasar phase occurs towards the end of this period,
just before the reservoir of cold gas is depleted.
Given the youth of the Universe at $z=4$ (1.6Gyr),
the coexistence of a massive black hole and a luminous starburst
at high redshifts is a powerful constraint
on models of quasar host-galaxy formation.

\end{abstract}


\section{INTRODUCTION}
Cosmological studies at FIR--submm wavelengths are an essential 
complement to optical investigations, in that the dust that
extinguishes the optical/UV light from AGN or massive stars
is that which thereby shines in the infrared.
This effect should not be underestimated: 
the integrated energy density in each of the FIR and optical
extragalactic background light is comparable (Puget \etal 1996;
Hauser \etal 1998).
It has been suggested that cosmological submm
sources are high-redshift equivalents of ULIRGs, their
luminosity deriving from bursts of
dust-obscured star-formation accompanying the assembly of 
massive elliptical galaxies,
in the course either of gas-rich hierarchical major-mergers 
or {\it ab initio} monolithic collapse (Barger, Cowie \& Sanders, 1999,
and references therein).
However, such speculations are difficult to confirm via
direct optical follow-up, because of
the large beam-sizes of current submm telescopes,
and because the dust obscuration necessarily makes the sources
optically faint.

An alternative approach is to study the submm properties of
well-characterised classes of objects, whose
precisely-known positions and redshifts, physical properties and 
population statistics, facilitate follow-up observations and interpretation.
To this end, we have been studying dust continuum and CO emission from
luminous ($M_B\la-27.5$), high-redshift ($z\ga4$) quasars.
At low redshifts, the hosts of luminous quasars are invariably massive
elliptical galaxies (Bahcall \etal, 1997; McLure \etal, 1999).
In the local Universe, massive black holes, fossil remnants of 
AGN now starved of fuel, are found in the cores of most galactic
bulges; moreover, their mass is correlated with the stellar mass
of the surrounding spheroid, 
$M_{\rm bh}=6\times10^{-3}M_{\rm sph}$ (Magorrian \etal, 1998). 
This correlation hints at 
a deep connection between black-hole build-up through accretion
and star-formation in the host galaxy.
Attributing the submm-luminosity of high-redshift quasars to 
star-formation therefore accords with the speculations about the
origins of the submm sources.

Before one can estimate physical quantities, 
such as dust temperature and star-formation rate, 
it is necessary to characterise the form of the 
dust-emission spectrum of these objects.
It is also important to quantify variations of the SED with object class,
luminosity or redshift, otherwise, if one adopts a single template for
all sources, it is not legitimate to adopt more than one derived parameter
as an indicator of physical status.
In fitting thermal SEDs, however, we are constrained
by the paucity of photometric data, and often it is inappropriate
to fit both the emissivity and the temperature as free parameters,
when we have only three or four photometric points per source.
Previous work, which considered object-by-object temperature fits,
adopted an assumed value for the emissivity index (Buffey \etal, in prep.).
In contrast, the purposes of this Letter are, first, to investigate 
the possibility of co-adding data from a sample of luminous, $z\ga4$ 
quasars, to find a self-consistent, overall best-fit 
temperature and emissivity; and to characterise the uncertainties
involved in doing so, whether they are due to experimental errors or
to physical differences. Finally, we discuss applications of the mean
SED, and try to justify the inferred parameters in terms of a simple
evolutionary model.

Throughout, we assume a $\Lambda$-dominated cosmology
$\Omega_M$=0.3, $\Omega_{\Lambda}$=0.7, 
$H_0$=65\kmps\Mpc$^{-1}$.

\section{MEAN ISOTHERMAL SPECTRUM}

\subsection{Data}
The $z>4$ datasets are from our work with IRAM (1.25\mm)
and JCMT (SCUBA 850\mic, 750\mic, 450\mic\ and 350\mic; UKT14 800\mic)
as described in companion papers 
(Isaak \etal 1994; Omont \etal, 1996a; Buffey \etal, in prep.);
and 350\mic\ CSO photometry is from Benford \etal (1999).
Additionally, we include the $z=3.91$ quasar APM08279$+$5255 
(Lewis \etal 1998).
All the sources have been detected at least four different
wavelengths. 
An important caveat
is that the quoted statistical uncertainties do not reflect 
the systematic errors due to flux calibration. 
Therefore, we assume calibration errors, combined in quadrature with the
random uncertainties, of 20 percent.

\begin{table}
\caption{Summary of SED parameters used by other authors}
\begin{tabular}{lcccc}
Authors & Sample &$\frac{\overline{L_{\rm fir}}}{\Lsun}$&$\overline{T}$
&$\overline{\beta}$\\ 
\hline
%Draine \& Lazarian & Galaxy & &20 & 1.7\\
Dunne \etal & SLUGS & 10$^{11}$ & 36 & 1.3 \\
Carico \etal & LIRGs & 10$^{11.5}$ & 32 & 1.6\\ 
Lisenfeld \etal & LIRGs & 10$^{11.5}$ & 38 & 1.6 \\
Benford \etal & High-$z$ objects & 10$^{13}$& 52 & 1.5$^{\dagger}$ \\
Hughes \etal 
& High-$z$ objects & 10$^{13}$& 50$^{\dagger}$ & 1.5$^{\dagger}$ \\
McMahon \etal & $z>4$ quasars & 10$^{13}$& 50$^{\dagger}$ & 1.5$^{\dagger}$ \\
This work & $z>4$ quasars & 10$^{13}$& 42 & 1.9 \\
\hline
\end{tabular}

$^{\dagger}$ values presumed, not determined%\\
\end{table}

\subsection{Assumptions}
The rest-frame far-infrared (FIR) SED can be described by a
modified blackbody spectrum, assuming that the grains are in 
equilibrium with the radiation field and that the dust cloud is 
optically-thin at submm wavelengths:
$S_{\nu} \propto \frac{\nu^{3+\beta}}{e^{\frac{h\nu}{kT}}-1}.$
%%%SENTENCE ADDED IN RESPONSE TO REFEREE'S COMMENT%%%%
Note, however, that some ultraluminous starbursts
(such as Arp220; Rowan-Robinson
\& Efstathiou, 1993) are optically thick at the shortest rest-wavelengths
sampled by the present data (60$\mu$m).  
%%%%%%%%%%%%%%%%%%%%%%%%%%%%%%%%%%%%%%%%%%%%%%%%%%%%%%
This method determines
the most effective {\it isothermal} parameterisation of the long-wavelength
emission, a simple, convenient description that is frequently adopted
(\eg Benford et al., 1999; Lisenfeld, Isaak \& Hills, 2000;
Dunne et al., 2000), though
in reality, a distribution of dust temperatures is likely. 
%%%Isothermality is perhaps not realistic, but it is nevertheless
%%%a convenient and oft-used parameterisation. 
%%%Sensitive mid-infrared data will be required to 
%%%discriminate the AGN contribution
Shorter wavelength radiation, in the near-to-mid infrared region 
($\sim$10\mic\ rest-frame),
is likely to be contaminated by the hotter dust that
absorbs the harder continuum due to
the AGN itself ({\it e.g.} Rowan-Robinson, 1995).
The contribution from star formation, on the other hand,
emerges in
%On the other hand, the star-formation contribution emerges in  
%the {\it cooler} components, due perhaps to star-formation rather than AGN, 
%dominate 
%emerge in 
the submillimetre region.
%emission.
To assess the contamination of the submillimetre luminosity due to
%contribution of 
the AGN component %to the submillimetre luminosity 
requires better mid-infrared constraints than exist for
all but a few high-redshift quasars (see, for example, Rowan-Robinson, 2000).
%Therefore here we focus merely on the simple, 
%convenient isothermal parameterisation.
%For the purpose of determining a star 
We therefore simply assume that the component 
extracted by our fitting procedure
is appropriate for the extrapolation of
%of the SED required to determine
a far-infared luminosity ($L_{\rm FIR}$) from which a star formation 
rate is to be calculated.
We caution, however, that the unforeseen
presence of a hot component could skew the 
fit to higher temperatures, leading us to overestimate the star formation
rate.

%c.f. SLUGS get skewed T because of isothermality
%APM08279 IRAS detections===>very hot, but...
%BR1202 ISOCAM===>no need for hot!

\subsection{Method}
Adopting $T$ and $\beta$ as free parameters, 
we employ a $\chi^2$-minimisation procedure to determine the
best single-component fit to the data, which is coadded in a 
self-consistent way. A subtlety arises because
for a given observed wavelength, the range of objects' redshifts
causes us to sample their spectra at different rest wavelengths.
Therefore, if we choose to normalise the data at a given
rest wavelength, a spectrum must be assumed so that an extrapolation
to this reference from the nearest {\it observed} data point can be performed.
The best-fit spectrum thus determined then itself becomes the template for 
the extrapolation, and the process is iterated.
In practice, this procedure converges to a solution rapidly 
({\it i.e.} without roaming around parameter space)
so for these datasets at least, we conclude that
the {\it overall} $\chi^2$-minimised fit is, indeed, the best self-consistent
solution.

\begin{figure}
%\begin{center}
%\psfig{figure=q+l+g_chis.ps,width=100.0mm,angle=0}
\epsfig{figure=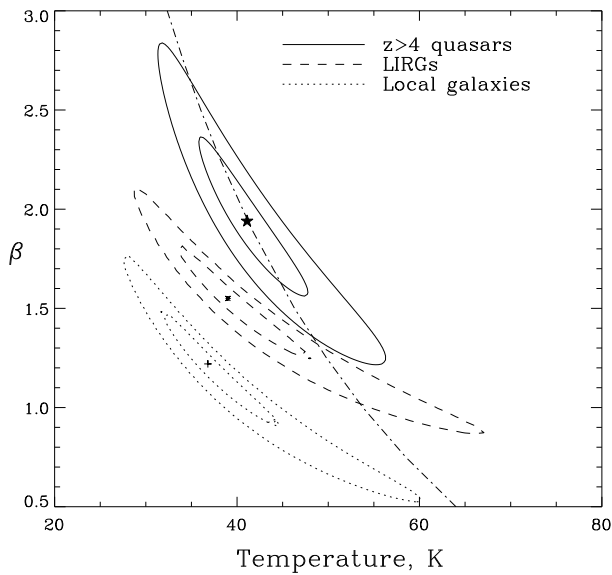,width=80.0mm,angle=0}
%\end{center}
\bigskip
\caption{Contours of $\chi^2$, in $\beta-T$ space,
for our combined photometry of $z>4$ quasars.
The star-shaped symbol indicates the best-fit position, and
both 1-- and 2--$\sigma$ contours are plotted (solid).
For comparison, the best-fit and 1-- and 2--$\sigma$ contours for local
LIRGs (from Lisenfeld, Isaak \& Hills, 1999) (asterisk and dashed),
and for local galaxies (from Dunne \etal, 2000) (cross and dotted),
are also shown.
The dot--dashed line is a locus of constant luminosity 
derived from the best-fit $T$ and $\beta$, for constant 850\mic\ flux
at $z=4.5$.
}
\end{figure}

\begin{figure}
%\begin{center}
%\psfig{figure=fig2.ps,width=100.0mm,angle=0}
\epsfig{figure=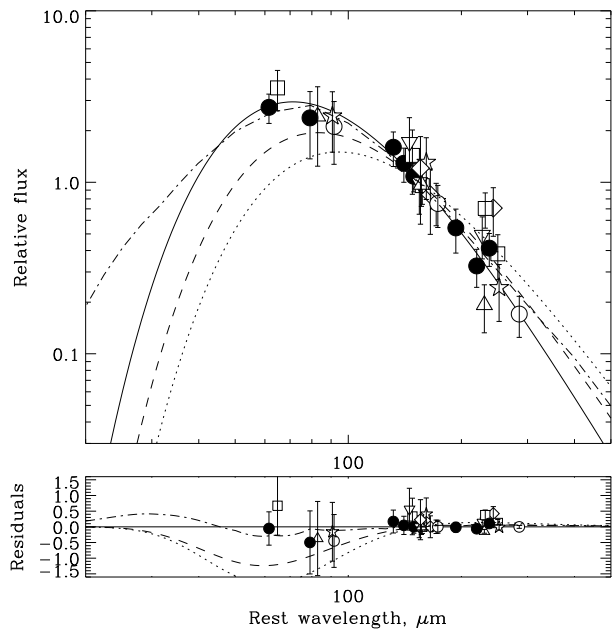,width=80.0mm,angle=0}
%\end{center}
\bigskip
\caption{
The best-fit, rest-frame SED, with renormalised
data points. Each separate symbol denotes a different quasar: 
BR1202$-$0725 (filled circles), APM08279$+$5255 (open circles),
BRI1335$-$0417 (squares), BR1117$-$1329 (stars), BRI0952$-$0115 
(upright triangles), BR1033$-$0327 (inverted triangles),
BR1144$-$0723 (diamonds).
The solid line corresponds to $T=41\K$, $\beta=1.95$, the $\chi^2$-minimised 
best fit to these points, while the dashed and dotted lines represent the
best fits to the data of Lisenfeld \etal and Dunne \etal (as in Figure 2). 
The dot--dashed
line is the starburst model of Rowan-Robinson \& Efstathiou (1993).
The curves are normalised at 850\mic/(1+4.5).}
\end{figure}

\begin{figure}
\epsfig{figure=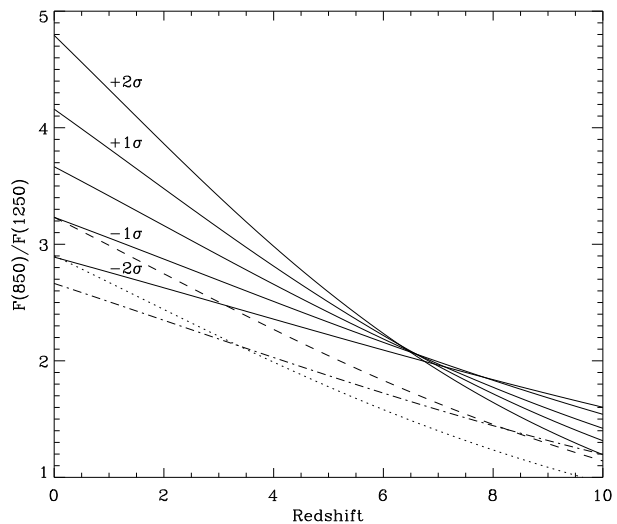,width=80.0mm}
\caption{Ratio of observed flux at 850\mic\ to that at 1250\mic---
\ie, the conversion between typical observations made with JCMT/SCUBA 
and with IRAM respectively. The solid lines are best fit (centre) and
$\pm$1-- and 2--$\sigma$ . The apparent convergence at $z\sim6$ is 
due to the tight bunch of data points near rest-frame 150\mic\
(see Figure 2). The dashed and dotted lines correspond, as in Figures 1 and
2, to the best-fits for the Lisenfeld \etal and Dunne \etal samples,
whilst the dot--dashed line depicts Arp220.
}
\end{figure}

\subsection{Results}

The $\chi^2$ contours are shown in Figure 1, while
the renormalised data and best-fit SED are plotted 
in Figure 2. The best self-consistent fit corresponds to 
$\beta=1.95\pm0.3$, $T=41\pm5\K$ (1$\sigma$ errors).
A similar procedure applied to the low-redshift, less-lumimous
samples of LIRGs (Lisenfeld, Isaak \& Hills, 1999) and
local galaxies (Dunne \etal, 2000) gives results inconsistent
at the $>2\sigma$ level (Figure 1). 
%It is worth noting that
%the emissivity is consistent with the value $\beta=1.7$,
%derived from the well-determined SED of cool Galactic ``cirrus'' dust
%observed by {\it COBE} (Draine, 1999). 

We can use the $\chi^2$ contours to predict the outcome of
curve-fitting in which $\beta$ is held fixed, while the temperature
is varied.
For other studies of high-redshift quasar
SEDs, the $\beta=1.5$ adopted
by Benford \etal (1999) and Buffey \etal (in prep.) would
give $T$=50\K\ as mean best-fit, which is indeed close to that
found by those authors, and consistent with the $T$=50\K\ assumed
by McMahon \etal (1999). 
Variation in 
luminosities (hence, eventually, star-formation rates) 
estimated from such best-fits do not vary substantially, because
$L_{\rm FIR} \propto (3+\beta)!\times T^{4+\beta}$,
whilst the best-fit $T$ is monotonically decreasing with $\beta$:
the dot--dashed line in Figure 1, a locus of constant luminosity, 
illustrates this.
This is to be expected, since luminosity is an integral under
the curve defined by the data points. 
On the other hand,
using best fits to the datasets consisting of other objects
(listed in Table 1), would give a discrepancy in luminosity of
up to a factor three at the redshifts of our quasars.

%Explicitly, this integral evaluates to
%
%
%\begin{equation}
%L(\beta,T) = 2\pi h c^2 {\left( \frac{k T}{h c}\right)}^{4+\beta}
%\Gamma(4+\beta)\zeta(4+\beta),
%\end{equation}
%
%\noindent where $\Gamma(x)=\int_{0}^{\infty}t^{x-1}e^{-t}dt$,
%which is simply $(x-1)$! when $x$ is an integer; 
%and $\zeta(x)=\sum_{k=1}^{\infty}k^{-x}$
%

%%%TABLE 2 USED TO BE HERE

\section{APPLICATIONS}
In this section, we discuss some applications for observation,
and implications for interpretation, of this mean SED.

\subsection{Millimetre--submillimetre spectral index}
The steep slope of cool, thermal SEDs over the mm--submm range
makes knowledge of the precise spectral shape important, 
when comparing data obtained with different instruments 
operating at different wavelengths. 
Consider, for example, the spectral index between 850\mic\ and
1.25mm, appropriate for converting between JCMT/SCUBA and IRAM data,
respectively.
The ratio of fluxes as a function of redshift for our best fit 
to $z>4$ quasars, together with
the fits to LIRGs and local galaxies, is shown in Figure 3.
For ease of reference, a linear approximation to the ratio for our best
fit, good out to $z\sim7$, is $R(\frac{850}{1250})\approx3.7-z/4$.

\begin{table*}
%\begin{center}
%\begin{minipage}{150mm}
\caption{$z>4$ sources used in the fit, from Omont \etal (1996),
with measured and extrapolated properties based upon the 850\mic\ fluxes
of Buffey \etal (in prep.)  }
\begin{tabular}{lccccccccccc}
Object & $z$ & $M_B$ & 
$M_{\rm bh}$ & $L_{\rm FIR}$ & $M_{\rm d}$ & $M_{\rm H_2}$ &
$M_{\rm sph}$ & $SFR$ & $\tau_{\rm sph}$ & $z_{\rm max}$ & $z_{\rm min}$\\
&&& 
$10^{9}\Msun$ & $10^{13}\Lsun$ & $10^9\Msun$ & $10^{11}\Msun$ &
$10^{11}\Msun$ & $10^{3}\frac{\Msun}{\rm yr}$ & $10^9\yr$ &&\\
\hline
BRI0952$-$0115& 4.43 &$-$28.0 & 5.5 & 2.0 & 1.7 & 1.4 & 
9.1 & 1.2 & 0.9 & 9.5 & 2.9\\
BR1202$-$0725 & 4.69 &$-$28.8 & 11  & 5.8 & 5.0 & 4.1 & 
19  & 3.5 & 0.7 & 7.9 & 3.3\\
BR1335$-$0417 & 4.40 &$-$27.6 & 3.8 & 2.0 & 1.7 & 4.3 & 
6.3 & 1.2 & 0.6 & 6.9 & 3.2\\
BR1033$-$0327 & 4.50 &$-$27.9 & 5.0 & 1.0 & 0.8 & $<$0.9$^{\dagger}$ &    
8.3 & 0.6 & 1.7 & $\infty$ & 2.3\\
BR1117$-$1329 & 3.96 &$-$28.2 & 6.6 & 1.9 & 1.7 & $<$1.2$^{\dagger}$ &
11 & 1.1 & 1.2  & 10 & 2.5\\
BR1144$-$0723 & 4.14 &$-$27.7 & 4.1 & 1.0 & 0.9 & $<$1.3$^{\dagger}$ &
6.9 & 0.6 & 1.4 & 20 & 2.4\\
\hline
Median Quasar$^{\ddagger}$&4.5 &$-$28.0 & 5 & 2 & 2 & 1.5 & 
8  & 1.2 & 0.7 & 7.4 & 3.2\\
\hline            
\end{tabular}
\begin{minipage}{170mm}
{\bf Note:} All derivations assume $\Omega_{M}=0.3$, $\Omega_{\Lambda}=0.7$,
$H_0=65\kmps\Mpc^{-1}$ \\
$^{\dagger}$ 3$\sigma$ upper limits\\
$^{\ddagger}$ represents median of the observed quasars, consistent with
an object having $S_{850}\approx15$mJy and $M_B=-28.0$.

\end{minipage}
%\end{center}
\end{table*}

\subsection{Physical characteristics}
Table 2 lists physical properties inferred for each of the quasars 
contributing to the fit. Dust masses ($M_{\rm d}$) 
and far-infrared luminosities ($L_{\rm FIR}$) are
calculated as in McMahon \etal (1999), though with our new $T$ and $\beta$.
The final stellar mass of the host spheroid ($M_{\rm sph}$) is estimated
using the local relation between bulge and black hole mass 
(Magorrian \etal, 1998), with the latter calculated from the absolute 
$B$-band magnitude ($M_B$) via simple assumptions about accretion
(Section 3.3), and a bolometric correction of 12 after Elvis \etal (1994). 
Molecular gas masses ($M_{\rm H_2}$) are calculated from observed
CO emission-line luminosities (Omont \etal 1996b; Guilloteau \etal 1997, 1998),
using a conversion factor 
$M_{\rm H_2}/L'_{\rm CO}=4.5\Msun (\K\kmps\pc^2)^{-1}$ 
(Solomon, Downes \& Radford, 1992); though we caution that a calibration as
much as five times smaller than this has been reported to be more 
appropriate for some ULIRGs (Downes \& Solomon, 1998).

The substantial star-formation rates and far-infrared luminosities might
cause us some concern: $L_{\rm fir}\sim L_B$, so we should consider
whether the AGN itself is an important contributor to the cool dust 
luminosity. Of course, the equally-substantial quantities of 
of cool molecular gas ($M_{H_2}>10^{11}\Msun$) make
ideal nests for hatching young stars; and stars are required to 
manufacture the large mass of dust that is observed
($M_{\rm d}>10^9\Msun$).
Moreover, the CO(5--4)/1.35mm-continuum of BR1202$-$0725 
has been resolved into two components (Omont \etal, 1996), only one
of which coincides with the quasar, the other showing no sign of
AGN activity. Yun \etal (2000) have detected the 
three CO-luminous quasars at 1.4 {\thinspace GHz}: the reported fluxes are 
consistent with synchrotron emission from supernova remnants, 
and imply star-formation rates $\sim$1000\Msunpyr, 
in agreement with the submm predictions (Section 3.3, and Table 2).
Therefore, although we do not exclude the possibility that some of the
submm flux is due to AGN-heating, here we shall explore the hypothesis that
it derives wholly from star-formation.

This leaves us with somewhat extraordinary objects, which are not
only undergoing intense bursts of dust-obscured star formation, 
but simultaneously hosting powerful quasars. 
The age of the Universe at $z=4$ is 
1.6\Gyr: this lack of time imposes severe constraints
upon the processes governing both central black-hole growth (Turner 1993)
and the chemical evolution necessary to supply the large observed dust
masses (Edmunds \& Eales 1998). 
The statistics are, currently, very poor (only three $z>4$ objects
with CO-detections, biased towards the most luminous objects), 
nevertheless we assume that the Median Quasar of Table 2 
epitomises its class, to guide our quest for a consistent hypothesis.

\subsection{Quasar and host galaxy coevolution}
The far-infrared luminosity ($L_{\rm FIR}$) can be used to estimate the 
current star-formation rate, given a stellar Initial Mass Function (IMF) 
and a prior star-formation history: 
$SFR/\Msunpyr=\Psi L_{\rm FIR}/10^{10}\Lsun$.
To estimate $\Psi$, we use the stellar population synthesis model
of Leitherer \etal (1999), {\tiny STARBURST99}, assuming a constant
star-formation rate and that $L_{\rm FIR}=L_{\rm bol}$.
A significant cause of uncertainty is due to the choice of IMF:
introducing faint, low-mass stars has the effect of increasing the total
mass while hardly affecting the luminosity, therefore one must infer
a higher star-formation rate for a given $L_{\rm bol}$.
We adopt a Salpeter IMF ($\phi(m)\propto m^{-2.35}$, Salpeter 1955) 
within the mass
range $1\Msun<m<100\Msun$.
$\Psi$ depends much more weakly upon metallicity, and upon the age
of the burst for lifetimes longer than $\sim10^7\yr$. 
We take $\Psi=0.6$, which corresponds to a metallicity
$Z=Z_{\odot}$ and an age $t\approx0.5\Gyr$.
%%%%SENTENCE ADDED IN RESPONSE TO REFEREE'S COMMENT%%%%
Another source of uncertainty is the fraction of starlight that
is intercepted by dust; here we make the simplifying assumption 
that this fraction is unity.
%%%%%%%%%%%%%%%%%%%%%%%%%%%%%%%%%%%%%%%%%%%%%%%%%%%%%%%
Star-formation rates thus derived are presented in Table 2.

The spheroid-formation timescales in Table 2 correspond to the
total time required to convert the inferred spheroid masses from gas 
into stars, at constant star-formation rate;
not forgetting gaseous material expelled, from massive stars,
as winds and supernovae (we assume a return fraction of $\approx$20 percent).
It is through this outflow process that the interstellar medium (ISM) becomes
chemically enriched: if half the mass in metals condenses into dust,
then chemical yields derived from stellar evolution models (\eg Maeder, 1992) 
predict that $\sim$1 percent of the mass processed through stars 
(for our IMF) is returned to the ISM in the form of dust---
neglecting however the intricacies of dust-destruction and -dispersion by
supernovae.
Thus, after 0.5\Gyr, the stellar mass of the Median Quasar would be 
$\approx$6$\times10^{11}\Msun$, its gas mass is $\approx$2$\times10^{11}\Msun$,
and its dust mass is $\approx$3$\times10^9\Msun$--- near the maximum 
it can attain for these parameters, before becoming recycled into
long-lived stellar remnants (Edmunds \& Eales, 1998).
%Moreover, we should ensure that the final dust-to-gas 
%ratio ($\Leftrightarrow$metallicity) is near that 
%observed for the local Universe 
%($\approx6\times10^{-3}$, $Z\approx Z_{\odot}$). 

Meanwhile, to describe the evolution of the AGN,
we adopt the simplest assumptions 
about accretion onto a massive black hole (Rees, 1984):
that the mass flow $\dot{M}_{\rm acc}$ is regulated by radiation pressure---
\ie accretion occurs at the Eddington rate--- and that the 
accretion efficiency $\epsilon\sim0.1$, so
\begin{equation}
L_{\rm bol}=\epsilon\dot{M}_{\rm acc}c^2=
\frac{4\pi G m_{\rm p} c M_{\rm bh}}{\sigma_{\rm T}}
=3\times10^{13}\frac{M_{\rm bh}}{10^9\Msun}\Lsun,
\end{equation}
where $\sigma_{\rm T}=6.65\times10^{-25}\cm^2$ 
is the Thomson cross-section, and 
$m_{\rm p}=1.67\times10^{-24}\g$ is the proton mass.
The accretion rate is
%\begin{equation} 
$\dot{M}_{\rm acc}=\frac{22}{\epsilon/0.1}
\frac{M_{\rm bh}}{10^9\Msun}\Msunpyr$
%\sim10^{-11}(\epsilon/0.1)^{-1}L_B\Msunpyr
%\end{equation}
and the mass $e$-folding time for the growth of the black hole is
$\tau_{\rm bh}=\frac{\epsilon M_{\rm bh} c^2}{L_{\rm bol}}=
\frac{\epsilon}{0.1}\times4.5\times10^7${\thinspace yr}. 
Supposing that Eddington-limited accretion starts to apply above
a seed black hole mass of about $10^6\Msun$ (Haehnelt, Natarajan \& Rees,
1998),
it would take $\sim$0.5{\thinspace Gyr} to build the required quasars--- 
close to the star-forming lifetime of the host galaxy,
discussed above.
%Even in this Lambda-dominated cosmology, these timescales encroach
%upon the maximum time available since the Big Bang. 
%However, adopting, for example, an exponentially-declining star-formation
%law and super-Eddington accretion would help to alleviate this.

The {\it total} (baryonic) mass of the galaxy 
has been inferred from the black hole mass by
taking the local {\it remnant} black hole--bulge correlation 
(Magorrian \etal. 1998).
%This is open to criticism on the grounds that 
Critics may object that we do not measure
the {\it final} mass of the black hole, because we observe it while it is
still actively accreting. However, by the time the black hole's mass is 
$5\times10^9\Msun$, its Eddington accretion rate is
100\Msunpyr and rising, and it is competing with star-formation 
($\sim$1000\Msunpyr) for the rapidly-diminishing supply of cold gas.
In the absence of inflow, this cannot be sustained for long.
%Since $\dot{M}_*\sim10^{-10}L_{\rm FIR}$, then
%$L_{\rm FIR}\sim L_B\Rightarrow\dot{M}_{\rm acc}\sim0.1\dot{M}_*$. 
Therefore, we obtain one consistent scheme by 
assuming that the black holes' masses do not grow much greater than 
the values deduced from the quasar light.
This is in accord with the measurements of remnant black hole masses in
the local Universe, which give at most $10^{9.5}\Msun$. 
%({\it e.g.} Ferrarese \& Merritt, 2000).
%This is in accord with estimates of quasars' {\it optical} lifetimes,
%which tend to be short, 
%a few $\times\tau_{\rm bh}$ at most (Haiman \& Hui, 2000).
%and ensures that estimates of remnant masses do not become prohibitive.

The quantities $z_{\rm max}$ and $z_{\rm min}$ in Table 2 bracket the
redshift range over which the quasars are forming stars, based upon 
the inferred spheroid mass and constant star-formation rate.
The severity of the time constraint at high redshifts 
is dramatically illustrated, especially for
BR1033$-$0327, which has a star-forming lifetime longer than the
age of the Universe at its redshift.
%This dramatically emphasises the importance of studying astrophysical
%phenomena at ever higher redshifts, where the constraints on one's models
%become prohibitive.
In this case, we could resolve the dilemma by invoking a 
different IMF and star-formation/accretion histories: 
for example,
exponential gas-consumption and a phase of
super-Eddington accretion, 
or more low-mass stars leading to a higher star-formation rate,
would give rather shorter lifetimes.
We refrain from investigating this here, however, for lack of space and
data-volume, but we shall examine the problem in later papers.

\section{CONCLUSIONS}

We have constructed a mean SED, over the rest-frame wavelength range
60--300\mic, for luminous $z>4$ quasars, assuming an isothermal 
dust spectrum. We consider applications of this SED, such as the
conversion between commonly-used submm/mm wavebands, and
the estimation of star-formation rates and dust masses. 
Using the observed local remnant black hole
to bulge mass relation to estimate the initial gas mass, we deduce 
star-formation timescales and sketch a simple evolutionary scenario.

Estimation of star-formation rate and dust and gas masses obtained
through submm--mm observations are potentially important constraints on models
of quasar--host-galaxy coevolution. However, the correct interpretation of
such information requires detailed knowledge of the SED and its variation with
parameters such as redshift and luminosity. Otherwise, it is misleading 
to treat, as independent, quantities whose derivation follows from a single
data point.
Although we have here determined a ``mean'' SED, we do not suggest that
all high-redshift quasars possess a unique dust temperature:
much of the scatter about the mean is very likely real.
The data are not yet extensive enough to justify a more detailed treatment,
particularly since only the most luminous objects are suitable for
multi-wavelength photometry.

We emphasise, too, that it is important to justify the large derived
star-formation rates and dust masses in terms of the evolution
of stellar populations and chemical abundances:
the simple treatments frequently adopted are not necessarily misleading, 
but neither are they immune to justified criticism.
Employing a skeletal evolutionary scheme,
a consistent account emerges, within the many uncertainties,
if we identify submm-luminous, high-redshift quasars
with an early, gas-rich phase in the formation of massive 
galaxies, at about the time when 
the dust mass reaches its maximum, after the AGN has grown sufficiently
luminous for it to be identified as a bright quasar and 
just before star-formation 
has depleted the fuel-supply upon which it subsists. 
We estimate that this quasar--starburst phase is short, 
lasting for little longer than 0.5{\thinspace Gyr}, though
by changing our assumptions--- for example, regarding
the IMF or early accretion and star-formation history---,
further consistent models could be constructed. 
Forthcoming observational results (Isaak \etal, in prep.)
will improve statistics and permit refinement
of these speculations.

\section*{ACKNOWLEDGMENTS}
RSP thanks PPARC for support, and RGM thanks the Royal Society.
We thank the referee for comments.


\begin{thebibliography}{}
\bibitem[\protect\citename{Bahcall \etal }1997]{bahcall}
Bahcall J.N., Kirhakos S., Saxe D.H., Schneider D.P., 1997,
ApJ, 479, 642

%\bibitem[\protect\citename{Barger \etal }1998]{barger}
%Barger A.J., Cowie L.L., Sanders D.B., Fulton E., Taniguchi Y., Sato Y.,
%Kawara K., Okuda H., 1998, Nature, 394, 248 

\bibitem[\protect\citename{Barger, Cowie \& Sanders }1999]{barger}
Barger A.J., Cowie L.L., Sanders D.B., 1999, ApJ, 518, L5

\bibitem[\protect\citename{Downes \& Solomon }1998]{ds}
Downes D. \& Solomon P.M., 1998, ApJ, 507, 615

\bibitem[\protect\citename{Dunne \etal }2000]{dunne}
Dunne L., Eales S., Edmunds M., Ivison R., Alexander P.,
Clements D.L., 2000, MNRAS, 315, 115

\bibitem[\protect\citename{Edmunds \& Eales}1998]{ee98}
Edmunds M.G, Eales S.A., 1998, MNRAS 299, L29

\bibitem[\protect\citename{Elvis \etal }1994]{elvis}
Elvis M. et al., 1994, ApJS, 95, 1

%\bibitem[\protect\citename{Ferrarese \& Merritt}2000]{fm}
%Ferrarese L. \& Merritt D., 2000, ApJL, 539, L9

\bibitem[\protect\citename{Guilloteau \etal }1997]{Guilloteau97}
Guilloteau S., Omont A., McMahon R.G., Cox P., Petitjean P., 1997, 
A\&A, 328, L1

\bibitem[\protect\citename{Guilloteau \etal }1997]{Guilloteau99}
Guilloteau S., Omont A., Cox P., McMahon R.G., Petitjean P., 1999, 
A\&A, 349, 363

\bibitem[\protect\citename{Haehnelt, Natarajan \& Rees}1998]{hnr}
Haehnelt M.G., Natarajan P., Rees M.J., 1998, MNRAS, 300, 817

\bibitem[\protect\citename{Hauser \etal }1998]{hauser}
Hauser M.G. et al., 1998, ApJ, 508, 25

\bibitem[\protect\citename{Hughes, Dunlop \& Rawlings }1997]{hdr}
Hughes D.H., Dunlop J.S., Rawlings S., 1997, MNRAS, 289, 766

\bibitem[\protect\citename{Isaak et al.\ }1994]{isaak}
Isaak K.G., McMahon R.G., Hills R.E., Withington S., 1994, MNRAS, 269, L28

\bibitem[\protect\citename{Leitherer \etal }1999]{leitherer}
Leitherer C. \etal, 1999, ApJS, 123, 3

\bibitem[\protect\citename{Lewis \etal }1998]{lewis}
Lewis G.F., Chapman S.C., Ibata R.A., Irwin M.J., Totten E.J.,
1998, ApJL, 505, 1

\bibitem[\protect\citename{Lisenfeld, Isaak \& Hills }2000]{lisenfeld}
Lisenfeld U., Isaak K.G., Hills R., 2000, MNRAS, 312, 433

\bibitem[\protect\citename{Maeder}1992]{maeder}
Maeder A., 1992, A\&A, 264, 105

\bibitem[\protect\citename{Magorrian \etal }1998]{magorrian}
Magorrian J. \etal, 1998, AJ, 115, 2285
  
\bibitem[\protect\citename{McLure et al.\ }1999]{mclure}
McLure R.J., Kukula M.J, Dunlop J.S., Baum S.A., O'Dea C.P.,
Hughes D.H., 1999, MNRAS, 308, 377

\bibitem[\protect\citename{McMahon et al.\ }1994]{mcmahon}
McMahon R.G., Omont A., Bergeron J., Kreysa E., Haslam C.G.T., 1994, 
MNRAS, 267, L9

\bibitem[\protect\citename{McMahon et al.\ }1999]{mcmahon99}
McMahon R.G., Priddey R.S., Omont A., Snellen I., Withington S.,
1999, MNRAS 309, L1

\bibitem[\protect\citename{Omont et al.\ }1996a]{omonta}
Omont A., McMahon R.G., Cox P., Kreysa E., Bergeron J., Pajot F., 
Storrie-Lombardi L.J., 1996a, A\&A, 315, 1

\bibitem[\protect\citename{Omont et al.\ }1996b]{omontb} 
Omont A., Petitjean P., Guilloteau S., McMahon R.G., Solomon P.M., 
P\'{e}contal, E., 1996b, Nature, 382, 428

\bibitem[\protect\citename{Puget \etal }1996]{puget} 
Puget J.-L. et al., 1996, A\&A, 308, L5

\bibitem[\protect\citename{Rees}1984]{rees84} 
Rees M.J., 1984, ARA\&A, 22, 471

\bibitem[\protect\citename{Rowan-Robinson}1995]{rr95} 
Rowan-Robinson M., 1995, MNRAS, 272, 737

\bibitem[\protect\citename{Rowan-Robinson}2000]{rr00} 
Rowan-Robinson M., 2000, MNRAS, 316, 885

\bibitem[\protect\citename{Rowan-Robinson \& Efstathiou}1993]{rre} 
Rowan-Robinson M. \& Efstathiou A., 1993, MNRAS, 263, 675

\bibitem[\protect\citename{Salpeter}1955]{salpeter} 
Salpeter E.E., 1955, ApJ, 121, 161

\bibitem[\protect\citename{Yun \etal }2000]{yun} 
Yun M.S., Carilli C.L., Kawabe R., Tutui Y., Kohno K., Ohta K.,
2000, ApJ, 528, 171

\end{thebibliography}
\end{document}